\documentclass [a4paper, twoside, french, 11pt]{article}
\usepackage[T1]{fontenc}
\usepackage[francais]{babel}
\usepackage{epsfig}
\usepackage{floatflt}
\usepackage{times}
\usepackage{amsfonts}
\input{epsf}

\newcommand{\B}{{\sf B}}	
\newcommand{\ab}{{\it AtelierB}}	

\newcommand{\Be}{\B\ \'e\-v\'e\-ne\-men\-tiel}

\newcommand{\refines}{{\sc refines}}	

\newcommand{\mach}{{\sc machine}}	
\newcommand{\sets}{{\sc sets}}	
\newcommand{\cons}{{\sc constants}}	
\newcommand{\prop}{{\sc properties}}	
\newcommand{\var}{{\sc variables}}	
\newcommand{\inv}{{\sc invariant}}	
\newcommand{\init}{{\sc initialisation}}
\newcommand{\oper}{{\sc operations}}	
\newcommand{\refi}{{\sc refinement}}	
\newcommand{\bend}{{\sc end}}		

\newcommand{\bif}{{\sc if }}	
\newcommand{\bthen}{{\sc then}}	
\newcommand{\bselect}{{\sc select }}	

\newcommand{\logand}{~\wedge~}		
\newcommand{\bdef}{~\hat =~}		

\newcommand{\nat}{\mathsf{NAT}}	
\newcommand{\true}{\mathsf{true}}	

\newcommand{\bob}{{\bf Bo\B}}

\newcommand{\GS}{{\sf G\'en\'eSyst}}	

\newcommand{\uml}{{\bf UML}}

\newcommand{\ste}{sys\-t\`e\-me de tran\-si\-tions \'e\-ti\-que\-t\'ees}
\newcommand{\stes}{sys\-t\`e\-mes de tran\-si\-tions \'e\-ti\-que\-t\'ees}

\begin{document}

 \title{\vspace*{-10pt}\itshape  \GS\ : G\'en\'eration d'un Système de transitions \'etiquet\'ees \`a partir d'une sp\'ecification \Be \vskip -10pt }
\author{Xavier Morselli {\footnotesize (xavier.morselli@imag.fr)} \and Marie-Laure Potet {\footnotesize (marie-laure.potet@imag.fr)} \and Nicolas Stouls {\footnotesize (nicolas.stouls@imag.fr)\footnote{{\scriptsize Ce travail est support\'e par une bourse BDI confinanc\'ee par le CNRS et ST Microelectronics.}}}\\
        {\footnotesize LSR-IMAG - 681, rue de la Passerelle,  BP72, 38402 St Martin d'H\`eres Cedex}}

\date{\vskip -10pt {\small Le \today }}

\footnotetext{\footnotesize Article publié dans les actes de la conférence AFADL'04 (http://lifc.univ-fcomte.fr/afadl2004/)}
  
\maketitle
\pagestyle{empty}
\thispagestyle{empty}

\centerline{{\bf R\'esum\'e}}

\vskip 7pt
\centerline{
 \parbox{0.9\textwidth}{
   \footnotesize
   La source d'erreur la plus couteuse et la plus d\'elicate \`a d\'etecter dans 
   un d\'eveloppement formel est l'erreur de sp\'ecification. 
   Ainsi la première phase d'un d\'eveloppement formel consiste 
   g\'en\'eralement \`a repr\'esenter l'ensemble des comportements possibles 
   sous la forme d'un automate.
   Partant de cette constatation, de nombreuses 
   recherches portent sur la g\'en\'eration d'une machine \B\ \`a partir d'une 
   sp\'ecification telle qu'\uml . \\ 
   \\
   Cependant, il faut ensuite \^etre capable de v\'erifier que la sp\'ecification 
   respecte bien les comportements d\'ecrits. C'est pourquoi, nous avons r\'ealis\'e un 
   outil, \GS , permettant d'extraire le contr\^ole d'un syst\`eme \Be\ et de le 
   repr\'esenter sous forme de \ste . 
   Le cas du raffinement est pris en compte et apparait sur l'automate produit 
   sous la forme d'\'etats hi\'erarchis\'es.\\
   \\
   Cet outil est une impl\'ementation des th\'eories d\'evelopp\'ees dans \cite{NSMLP04}.\\
   \vskip 3pt
   \textbf{Mots clefs :} M\'ethode \B , sp\'ecification, raffinement, syst\`emes de transitions.
 }
}

  \section{Fonctionnalités de GénéSyst}

   L'approche \Be\ permet de modéliser à la fois les données, leur 
   traitement et la dynamique d'un système. Cette approche est basée sur la 
   notion d'événements. Ceux-ci sont caractérisés par une garde (une condition de déclenchabilité) 
   et une action. La modélisation événementielle introduit une difficulté 
   supplémentaire : le raisonnement sur la dynamique du système. L'outil
   \GS\ a pour but de permettre de visualiser  cet aspect sous la forme
   de systèmes de transitions symboliques. Il permet de :
   \begin{itemize}
     \item représenter par un état un ensemble de valeurs (potentiellement infinis).
           
     \item \vskip 2pt calculer des transitions  conditionnées représentant les hypothèses 
           sous lesquelles un événement peut être déclenché (condition de déclenchabilité)
           dans un état donné et les hypothèses sous lesquelles un événement permet
           d'atteindre un état (condition d'atteignabilité).

     \item \vskip 2pt représenter le processus de raffinement par des systèmes de transitions
           hiérarchisés.
   \end{itemize}
   
   \vskip 2pt 
   Le premier point permet de visualiser de manière lisible des systèmes 
   manipulant des données appartenant à des domaines infinis ou grands.
   
   Le second point permet de préciser au mieux les conditions sous lesquelles
   une transition peut être activée. En particulier si elle n'est pas possible, 
   toujours possible ou conditionnée.
   
   Enfin, la représentation des raffinements par des systèmes de transitions
   hiérarchisés permet  de visualiser le lien entre deux niveaux de raffinement,
   en précisant la structuration du système de transitions abstrait.
   Le système de transitions obtenu est complet et correct vis-à-vis de la
   spécification : il admet exactement les mêmes traces que le système \Be\ \cite{NSMLP04}.

   Ces travaux ont été initialement introduits par D. Bert et 
   F. Cave \cite{BC00}. Ils ont ensuite été étendus par S. Hamdane \cite{Smaine}, 
   avant d'être complétés par N. Stouls et M-L Potet \cite{NSMLP04}.

\begin{figure}[h]
  \begin{center}
    \scriptsize
    \begin{tabular}{ll}
      \parbox{6cm}{
        \begin{tabbing}
          XXX\=XXX\=XXX\=XXX\=XXX\=XXX\=XXX\=XXX\=XXX\=XXX\=XXX\=XXX\=XXX\= \kill
          \mach \>\>\>~~$parking$\\
          \cons \>\>\>~~$NbPlaces$\\
          \prop \>\>\>~~$NbPlaces \in \nat \logand NbPlaces > 0$\\
          \var \>\>\>~~$NbVoit,cc$\\
          \inv \>\>\>~~$(NbVoit \!\in\! 0..NbPlaces)$\\
               \>\>\>~~$\!\logand\! (cc \!\in\! -1..1)$\\
               \>\>\>~~$\!\logand\! (cc\!=\!-1 ~\Rightarrow~ NbVoit\!<\!NbPlaces)$\\
               \>\>\>~~$\!\logand\! (cc\!=\!1 ~\Rightarrow~ NbVoit\!>\!0)$\\
          \init \>\>\>~~$NbVoit\!:=\!0~||~cc\!:=\!0$
        \end{tabbing}
        \vskip -24pt
      }
    & 
      \parbox{8.48cm}{
        \vskip -15pt
        \hskip -10pt
        \epsfxsize=8.48cm \epsfbox{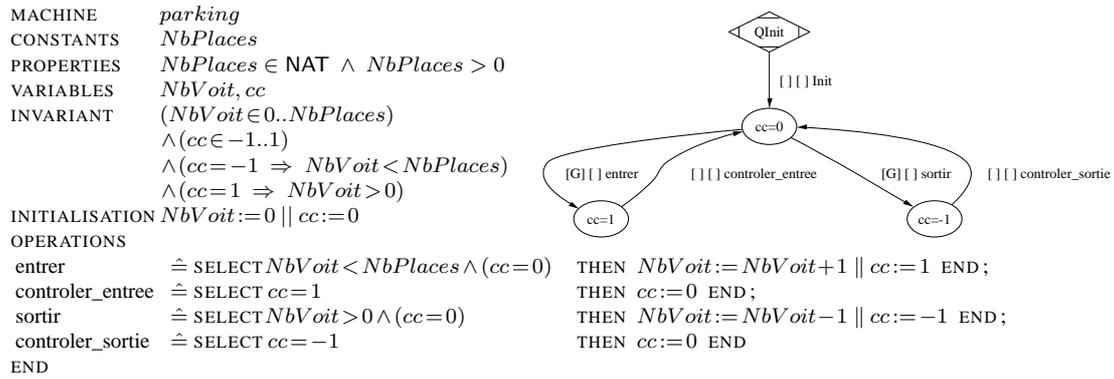} 
        \vskip -45pt
      }
    \\
      \multicolumn{2}{l}{
        \parbox{15cm}{
          \begin{tabbing}    
            XXX\=XXX\=XXX\=XXX\=XXX\=XXX\=XXX\=XXX\=XXX\=XXX\=XXX\=XXX\=XXX\= \kill
            \oper\\
            ~entrer\>\>\>~~~$\bdef $\bselect $\!NbVoit\!<\!NbPlaces \!\!\logand\!\! (cc\!=\!0)$ \>\>\>\>\>\>\>\>\>~~\bthen\ $~NbVoit\!:=\!NbVoit\!+\!1~\|~cc\!:=\!1~$ \bend ;\\
            ~controler\_entree\>\>\>~~~$\bdef $\bselect $cc\!=\!1$ \>\>\>\>\>\>\>\>\>~~\bthen\ $~cc \!:=\! 0 ~$ \bend ;\\
            ~sortir\>\>\>~~~$\bdef $\bselect $\!NbVoit\!>\!0 \!\!\logand\!\! (cc\!=\!0)$     \>\>\>\>\>\>\>\>\>~~\bthen\ $~NbVoit\!:=\!NbVoit\!-\!1~\|~cc\!:=\!-1~$ \bend ;\\
            ~controler\_sortie\>\>\>~~~$\bdef $\bselect $cc\!=\!-1$ \>\>\>\>\>\>\>\>\>~~\bthen\ $~cc \!:=\! 0 ~$ \bend\\
            \bend
          \end{tabbing}
        }
      }
    \end{tabular}
  \end{center}
  \vskip -30pt
  \caption{Exemple d'un parking avec contr\^oleur (inactif) et son système de transitions associé.}
  \vskip -5pt  
  \label{Parking2.ref}
\end{figure}

      La figure \ref{Parking2.ref} d\'ecrit un parking ayant un nombre de place limité ($NbPlaces$) et dans lequel 
   les véhicules peuvent $entrer$ ou $sortir$. Un contr\^oleur, pour le moment inactif, réagit à chacun 
   de ces 2 stimulis externes.
   Sur les \stes\ produits, une condition est not\'ee $[~]$ si elle est r\'eductible \`a $\true$ et est notée $[G]$ sinon.
   Ainsi, Les transitions $controler\_entree$ et $controler\_sortie$ ne sont pas conditionnées car $cc$ est instanci\'ee.
   En revanche, $entrer$ et $sortir$ d\'ependent de $NbVoit$ et sont donc conditionn\'ees.

\vskip -15pt

\begin{figure}[h]
  \vskip -10pt
  \centerline{
  \hskip 5pt
  \begin{tabular}{cc}
      \parbox{8cm}{
        \scriptsize
        \begin{tabbing}
          X\=X\=X\=X\=X\=X\=X\=X\=X\=X\=X\=X\=X\= \kill
          \refi
          \>\>\>\>\>\>\>\>$parking\_r1$\\
          \refines
          \>\>\>\>\>\>\>\>$parking$\\
          \sets
          \>\>\>\>\>\>\>\> $Couleur\_feu = \{vert, rouge\}$\\
          \var
          \>\>\>\>\>\>\>\> $feu, NbVoit, cc$\\
          \inv
          \>\>\>\>\>\>\>\> $feu \in Couleur\_feu $\\
          $\logand ((cc=0 \!\logand\! feu=vert)  \Rightarrow NbVoit<NbPlaces)$\\
          $\logand ((cc=0 \!\logand\! feu=rouge)  \Rightarrow NbVoit=NbPlaces)$\\
          $\logand (cc=1 \Rightarrow feu=vert)$\\
          $\logand (cc=-1 \Rightarrow NbVoit<NbPlaces)$\\
          $\logand ((cc=-1 \!\logand\! feu=rouge)  \Rightarrow NbVoit=NbPlaces-1)$\\
          $\logand ((cc=-1 \!\logand\! feu=vert)  \Rightarrow NbVoit<NbPlaces-1)$\\
          \init \>\>\>\>\>\>\>\>\>\>$NbVoit:=0 \| feu:=vert \| cc:=0$
        \end{tabbing}
        \vskip -25pt
    } &
    \parbox{6.5cm}{
      \vskip -5pt
      \hskip -40pt
      \epsfxsize=8.5cm \epsfbox{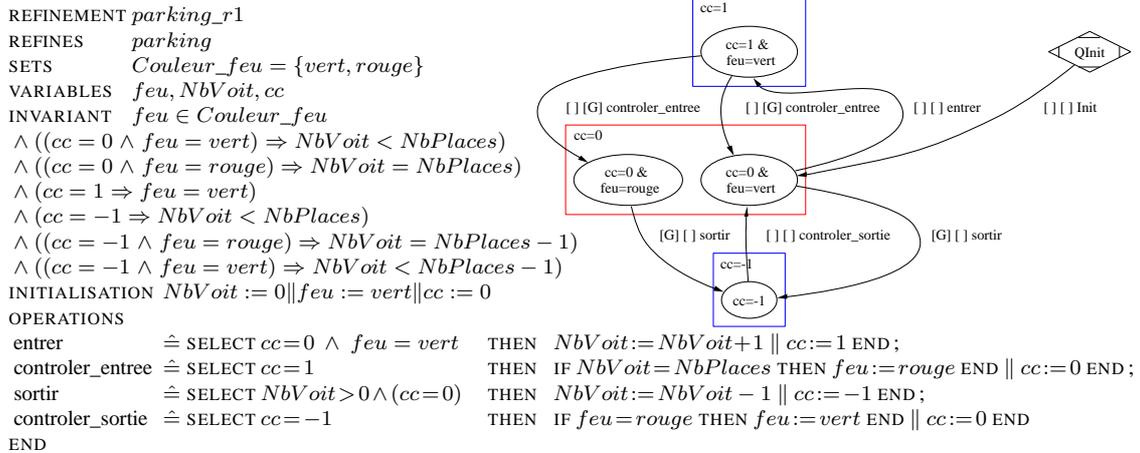}
      \vskip -23pt
    }\\ 
    \multicolumn{2}{l}{
      \parbox{16cm}{
        \scriptsize
        \begin{tabbing}
          XXX\=XXX\=XXX\=XXX\=XXX\=XXX\=XXX\=XXX\=XXX\=XXX\=XXX\=XXX\=XXX\= \kill
          \oper\\
          ~entrer            \>\>\>~~$\bdef $\bselect $cc\!=\!0 \logand feu=vert$ \>\>\>\>\>\>\>\>$\!\!\!\!\!\!\!\!$\bthen\ $~~NbVoit\!:=\!NbVoit\!+\!1~\|~cc\!:=\!1$ \bend ;\\
          ~controler\_entree \>\>\>~~$\bdef $\bselect $cc\!=\!1$ \>\>\>\>\>\>\>\>$\!\!\!\!\!\!\!\!$\bthen\ $~~$\bif $NbVoit\!=\!NbPlaces$ \bthen\ $feu\!:=\!rouge$ \bend\ $\|~ cc\!:=\!0 $ \bend ;\\
          ~sortir            \>\>\>~~$\bdef $\bselect $NbVoit\!>\!0 \!\!\logand\!\! (cc\!=\!0)$ \>\>\>\>\>\>\>\>$\!\!\!\!\!\!\!\!$\bthen\ $~~NbVoit\!:=\!NbVoit-1~\|~cc\!:=\!-1$ \bend ;\\
          ~controler\_sortie \>\>\>~~$\bdef $\bselect $cc\!=\!-1$ \>\>\>\>\>\>\>\>$\!\!\!\!\!\!\!\!$\bthen\ $~~$\bif $feu\!=\!rouge$ \bthen\ $feu\!:=\!vert$ \bend\ $\|~ cc\!:=\!0 $ \bend\\
          \bend
        \end{tabbing}
      }
    }
  \end{tabular}
  }
  \vskip -20pt
  \caption{Exemple du parking raffiné et son syst\`eme de transitions associ\'e}
  \vskip -5pt
  \label{Parking3.ref}
  \label{Parking3.ste}
\end{figure}

   La figure \ref{Parking3.ref}, d\'ecrit un raffinement du parking de la figure \ref{Parking2.ref}, 
   dans lequel un feu d'entr\'ee a été introduit. La gestion de celui-ci est effectuée 
   par le contr\^oleur. 

   \label{ParParking3.ref}

  \section{Réalisation}

   \GS\ prend en entrée un système \B\ décrit par une machine 
   ou un raffinement. L'utilisateur fournit les états du
   système de transitions sous la forme d'une disjonction  donnant
   les prédicats associés à ces états. Cette disjonction est fournie
   par l'intermédiaire de la clause ASSERTIONS. L'obligation de 
   preuve associée à cette clause va garantir que les états représentent
   toutes les valeurs possibles de l'invariant.

   Le résultat est fourni sous la forme d'un fichier représentant 
   le système de transitions. 
   Ce fichier, qualifié de {\it format intermédiaire}, est une représentation textuelle 
   du système de transitions produit. C'est à partir de celui-ci que \GS\ génère 
   des systèmes de transitions dans différents formats exploitables par d'autres outils (pour l'instant, seuls
   les formats {\it DOT} et {\it BCG} sont supportés).

\begin{floatingfigure}[l]{6cm}
     \centerline{\epsfysize=3cm \epsfbox{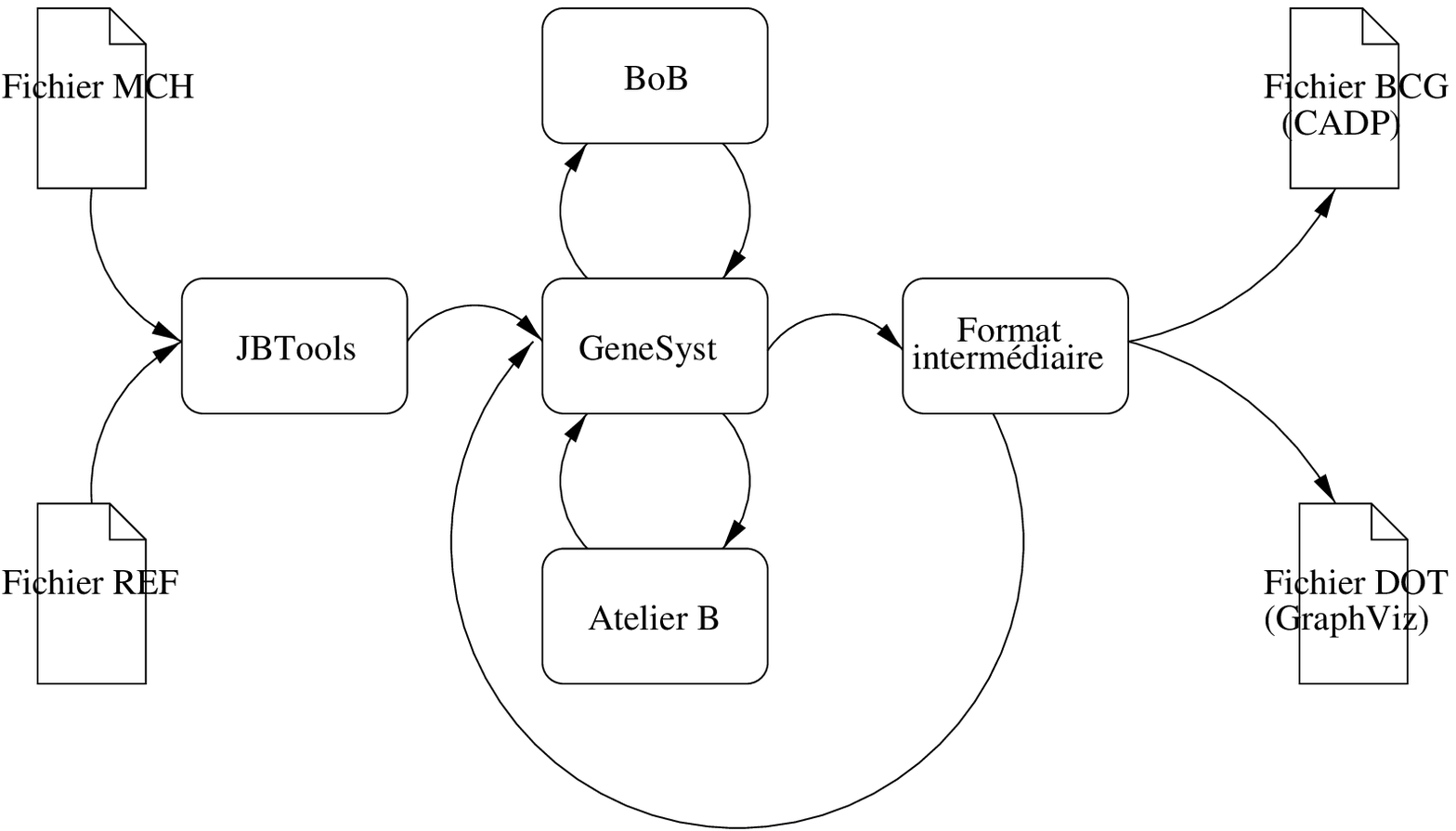}}
     \caption{Environnement de \GS }
     \label{InteractionsDeGS}
\end{floatingfigure}   

   \GS\ fonctionne en mode automatique. Pour chaque événement 
   il cherche à prouver si l'événement est toujours déclenchable 
   ou jamais déclenchable. Si une de ces preuves
   abouties, alors on a la réponse, sinon la transition est visualisée
   par défaut. Ceci ramène donc le problème de la recherche des
   conditions à un  problème de preuve. 
   Le même type d'approche est utilisée pour l'atteignabilité. 
   L'approche proposée et sa correction est décrite dans \cite{NSMLP04}.

   \GS\ est réalisé en Java en utilisant les outils suivants :
   \begin{itemize}
    \item le parseur JBTools \cite{jBTools} qui permet d'analyser les spécifications.
    \item la \bob\ (boite à Outils \B\ du LSR) qui permet de produire les obligations
           de preuve par calcul de plus faible pré-condition.
    \item le prouveur de l'\ab\  utilisé actuellement en mode automatique.
   \end{itemize}

  \section{Conclusions et perspective}
  
   \GS\ peut être utilisé en phase de mise au point d'une spécification
   ou d'un développement. Dans ce cadre, le système de transitions peut être
   modifié sans être remis en cause totalement. 
   \GS\ peut donc aussi utiliser en entrée une description (totale 
   ou partielle) d'un système de transitions. Dans ce cas son travail 
   consiste  à vérifier et corriger ce système de transitions.
   L'efficacité de l'outil est améliorée puisque le système de transitions
   permet de guider les preuves à faire.
   De plus, l'utilisateur peut ainsi simplifier les conditions sur les
   transitions et \GS\ produira les obligations de preuve assurant
   leur correction. Cette extension est en cours.

   Une autre amélioration consiste à  découpler l'activité de
   preuve de l'outil. Ceci permet soit d'affiner les preuves de manière
   interactive avec le prouveur de l'\ab\ 
   soit d'interfacer l'outil avec un autre démonstrateur.
   La qualité du résultat peut ainsi être améliorée puisque certaines
   transitions visualisées correspondent à un défaut de preuve automatique.

   D'un point de vue méthodologique, 
   l'introduction d'états hiérarchisés permet de décrire les 
   transitions à différents niveaux (entre les sur-états ou entre
   les sous-états). 
   Ceci nécessite d'élaborer des heuristiques permettant de faire
   ces choix de visualisation.
   Enfin, une interface graphique est en cours d'élaboration. Le but 
   est de pouvoir zoomer sur certaines parties du système de 
   transitions, par exemple pour voir le détail d'une transition ou
   d'un état.

   \GS\ est diffusée sur le site du LSR à l'adresse : 
   
   \centerline{http://www-lsr.imag.fr/Les.Personnes/Nicolas.Stouls/}

\bibliographystyle{alpha}
\bibliography{Bibliographie}

\end{document}